\documentclass[10pt]{article}
\usepackage{a4wide}
\usepackage{graphicx}
\usepackage{epstopdf}
\usepackage{amstext,amsmath,amsfonts,amsbsy,amssymb,bbm}
\usepackage{array,multirow}
\usepackage[pageshow]{supertabular}
\usepackage{times,color}
\usepackage{cite}
\newcommand {\be}[1]{\begin{eqnarray} \mbox{$\label{#1}$}  }
\newcommand{\ee}{\end{eqnarray}}

\newcommand{\pref}[1]{(\ref{#1})}

\newcommand{\nn}{\nonumber\\}
\newcommand{\noi}{\noindent}

\newcommand\hf{\frac 1 2 }
\newcommand\half{\frac 1 2 }
\newcommand{\com}[2]{\left[ #1, #2 \right]}

\newcommand{\dd}[2]{{d{#1}\over d{#2}}}
\newcommand{\ddtwo}[2]{{d^2{#1}\over d{#2}^2}}

\newcommand{\pdd}[2]{{\partial{#1}\over\partial{#2}}}

\newcommand{\pd}{\partial}

\newcommand{\gd}{ {\delta} }
\newcommand{\gD}{ {\Delta} }
\newcommand{\gk}{\kappa}
\newcommand{\gf}{\phi}
\newcommand{\gw}{ {\omega} }
\newcommand{\gq}{ {\theta} }
\newcommand{\gQ}{ {\Theta} }
\newcommand{\gy}{ {\psi} }
\newcommand{\gx}{ {\xi} }
\renewcommand{\ge}{ {\epsilon} }
\newcommand{\gl}{ {\lambda} }
\newcommand{\gr}{\rho}

\newcommand{\sgn}{{\rm sgn}}
\newcommand{\te}{{\tilde \ge}}
\newcommand{\tf}{{\tilde f}}
\newcommand{\brk}{{\bar k}}

\begin{document}
\centerline{\Large \bf Fermi liquids and fractional statistics in one dimension.\footnote{Contribution to the Frank Wilczek Festschrift on the occasion of his 70th birthday.}}
\vskip 15mm
\centerline{\bf Jon Magne Leinaas}
\medskip
\centerline{Department of Physics, University of Oslo}
\centerline{P.O. Box 1048 Blindern, N-0316 Oslo, Norway}

\begin{abstract}
Interacting fermion systems in one dimension, which in the low energy approximation are described by Luttinger liquid theory, can be reformulated as systems of weakly interacting particles with fractional exchange statistics. This is shown by use of Landau's Fermi liquid theory, with quasiparticles interpreted as adiabatically dressed fermions. An application of 
this method is included, where boundary excitations of a two-dimensional quantum Hall    electron system are studied.
\end{abstract}

\section{Introduction}

Ideas of quantum particles with unconventional statistics have over the years been met with much interest, both with respect to theory and  to experimental verifications. In three dimensions elementary particles are restricted to the two main types, fermions and bosons. However, in condensed matter systems, where the number of dimensions is effectively reduced to two or one, the elementary (particle-like) excitations can, under certain conditions, satisfy statistics that is different from the two elementary types. In particular, in the fractional quantum Hall effect, the electrons form collective states with new (quasi)particles that are characterized both by fractional statistics and fractional charges.

There are more than one type of unconventional statistics that has theoretically been introduced. The type of statistics seen in the quantum Hall effect, often referred to as {fractional {\em exchange} statistics, describes identical particles where the quantum wave function picks up a complex phase factor when the position of two particles are interchanged \cite{LeinaasMyrheim77}. Bosons and fermions are then only special cases, where the phase factor is respectively $+1$ and $-1$.
Particles with other values of the phase factor are generally called {\em anyons} \cite{Wilzcek81}.

Anyons are restricted to systems which effectively are two-dimensional. However, other types of generalized statistics are possible in one dimension. One possibility 
is based on the assumption that for systems of identical particles only symmetric combinations of particle operators should be regarded as observables \cite{LeinaasMyrheim88}. This restriction opens for new types of quantum many-particle states, and in one dimension the new possibilities form a one-parameter set of particle representations, which includes also here bosons and fermions as special cases.

A third suggestion of generalizing quantum statistics 
is different from the two other by introducing a generalization of the Pauli exclusion principle in many-particle systems, rather than modifying the definition of exchange symmetry of the particles
\cite{Haldane91}. 
Also this gives rise to possible generalizations of quantum statistics in one dimensional systems. This type of statistics is often referred to as fractional {\em exclusion} statistics. 

I will here show how fermion systems in one dimension, quite generally can be described in terms of particles with fractional exclusion statistics.  The starting point is a general fermion system with a fermi sea as ground state. A short description of the Luttinger liquid description of the system is introduced \cite{Haldane81}, and a Fermi 
liquid description is derived with {\textquoteleft dressed\textquoteright} fermions as Landau quasiparticles \cite{Landau56}. The Landau (interaction) parameters are identified, and by use of a functional transformation of the variables it is shown that the Landau parameters can be nullified, and that the new variables satisfy fractional exclusion statistics \cite{Leinaas17}. A concrete realization is then discussed, where a quantum Hall system with boundary excitations effectively forms of a one-dimensional system. The results concerning interactions between the two boundaries are discussed and compared with earlier evaluations of the effect. 

Much of the first part of the paper is a review of results from an earlier publication \cite{Leinaas17},  however here with less details in the derivation of the fractional statistics. Relations between exclusion statistics and Luttinger liquid \cite{WuYu95,WuYu01}, and between exclusion statistics and Fermi liquid theory \cite{Isakov98}, have been discussed before, but the approach that is discussed here is different. The convention $\hbar=1$ is used throughout the paper.

\section{The Luttinger liquid formulation}
The starting point is the following general expression for the Hamiltonian of a one-dimensional system of spinless fermions,
\be{ham1}
H=\sum_k\ge_0(k)c_k^\dag c_k+{1\over{4L}}\sum_{q,k_1,k_2}V(k_1-k_2,q)c_{k_1}^\dag c_{k_2}^\dag c_{k_2-q}c_{k_1+q}\,.
\ee
The dependence of $V$ on the momentum variable $k_1-k_2$, in addition to $q$, opens for the possibility of a non-locality of the interaction (see Sect.~\ref{seven}).
However, the assumption is that the dependence on this variable is weak. This implies that the effect of the $k$-dependence for low energy particles close to the same Fermi point can be disregarded, while for the interaction between particles at opposite Fermi points the effect may be significant.   
The one-dimensional fermion system is assumed to have a ground state in the form of a filled Fermi sea, with well-defined boundaries at the two Fermi points $k=\pm k_F$ in momentum space, and with gapless, low energy excitations. Under these assumptions the  Hamiltonian of the system, in the low-energy approximation, can be simplified by the linearized expression \cite{Haldane81},
\be{ham2}
H=  \bar v_F\sum_{\chi,k}(\chi k-k_F):c_{\chi,k}^\dag c_{\chi,k}: +{1\over{4L}}\sum_{\chi,q}[V_1(q)\gr_{\chi,q}\gr_{\chi,-q}+V_2(q)\gr_{\chi,q}\gr_{-\chi,-q}]\,.
\ee
$\chi=\pm 1$ is here a chirality parameter, associated with the two Fermi points, and $\gr_{\chi,q}=\sum_k :c_{\chi,k+q}^\dag c_{\chi,k}:$ are the Fourier components of the particle density of chirality $\chi$, normal ordered relative to the filled Fermi sea. The system is assumed to be confined to an interval of length $L$, with periodic boundary conditions for the fermion fields. ($L$ is assumed to be much larger than any  physical length, and may be taken to infinity at places where that is convenient.) The momentum then takes discrete values $k=2\pi n/L$, with $n$ as an integer. The interaction is separated in two parts, with $V_1(q)$ as the interaction between fermions with the same chirality and $V_2(q)$ with opposite chiralities.  The effective Fermi velocity $\bar v_F$ has the form
\be{Fermi}
\bar v_F=v_F-{1\over 4\pi }(V_1(0)-V_2(0))\,,
\ee
with $v_F=\pdd{\ge_0}{k}(k_F)$ as the Fermi velocity of the non-interacting Fermi system, and the interaction dependent term is a correction, created by interactions between the low energy fermions and the Fermi sea \cite{Horsdal07}.

Although the quantum number $k$ is, in the linearized approximation, restricted to small deviations from $\pm k_F$, this restriction can be lifted, since the low energy sector of the theory is not affected by this extension. Without the restriction, the model \pref{ham2} describes in effect two types of fermions, characterized by different values of  $\chi$, both types with linear dispersion. 

The standard way to analyze the system described by the Hamiltonian \pref{ham2} is in terms of bosonization \cite{Haldane81}. I briefly summarize expressions to be used in the discussion to follow. 
The Fourier components of the charge density operators, $q\neq 0$, define the boson annihilation and creation operators,
\be{aop}
a_q=\sqrt{{2\pi}\over |q|L}\sum_\chi \gq(\chi q)\,\rho_{\chi,\,q}\,,\quad 
a_q^\dag=\sqrt{{2\pi}\over |q|L}\sum_\chi \gq(\chi q)\,\rho_{\chi,\,-q}\,,
\ee
where $\gq(q)$ is the Heaviside step function, and the $q=0$ components of the charge densities define  the conserved fermion numbers 
\be{NJ}
N=\sum_{\chi}N_\chi=\sum_{k\, \chi}:c^\dag_{\chi,k} \,c_{\chi,k}:\,,\quad\quad J=\sum_{\chi}\chi N_\chi=\sum_{k\, \chi}\chi\, c^\dag_{\chi,k}\,c_{\chi,k}\,,
\ee
with $N$ measuring the deviation of the particle number from its ground state value $N_0$. The bosonized form of the Hamiltonian is \cite{Haldane81}
\be{boseham}
H&=&{{\pi }\over{2L}}(v_N N^2+v_JJ^2)
\nn
&+&{ 1\over 2}\sum_{q\neq 0}|q|\left[
(\bar v_F+\frac{V_{1}(q)}{4\pi })
(a_q^{\dag}a_q+a_q a_{q}^{\dag})+\frac{V_{2}(q)}{4\pi }(a_q^{\dag}a_{-q}^{\dag}+a_q a_{-q})\right] \,,
\ee
Compared to the Hamiltonian \pref{ham2} it is modified by removing (non-relevant) terms that are constant or linear in $N$.  
The two velocity parameters $v_N$ and $v_J$ are 
\be{velpar}
v_N&=&\bar v_F+\frac{1}{4\pi }(V_1(0)+V_2(0))= v_F+\frac{1}{2\pi }V_2(0)\,,\nn
v_J&=&\bar v_F+\frac{1}{4\pi }(V_1(0)-V_2(0))= v_F\,.
\ee
One should note that $v_J$ is identical to the original Fermi velocity $v_F$ of the non-interacting fermions, rather than to the effective Fermi velocity $\bar v_F$, which appears in the Luttinger Hamiltonian \pref{ham2}. 
This can be viewed as a consequence of Galilei invariance of the  Hamiltonian (see Sect.~\ref{four}).
The low energy sector, where \pref{boseham} is valid, corresponds to situations where $|q|$, as well as $N/L$ and $|J|/L$, are effectively restricted to values much smaller than $k_F$.

The bosonized Hamiltonian is diagonalized by a Bogoliubov transformation of the form
\be{bogo}
a_q=\cosh\xi_q\,b_q+\sinh\xi_q\,b^\dag_{-q}\,,\nn
a^\dag_q=\cosh\xi_q\,b^\dag_q+\sinh\xi_q\,b_{-q}\,,
\ee
where $\xi_q$ is fixed by the relation
\be{xi}
\tanh2\xi_q=-\frac{V_2(q)}{V_1(q)+4\pi  \,\bar v_F}\,.
\ee
In terms of the new bosonic operators the Hamiltonian gets the diagonal form
\be{hamdiag}
H=\sum_{q\neq 0} \gw_q \,b_q^{\dag}b_q
+ {{\pi }\over{2L}}(v_N N^2+v_JJ^2)\,,
\ee
with the frequency $\gw_q$ given by
\be{freq}
\gw_q=\sqrt{\left(\bar v_F+\frac{V_1(q)}{4\pi }\right)^2-\left(\frac{V_2(q)}{4\pi }\right)^2}\;|q|\,.
\ee

The bosonized form of the low-energy Hamiltonian \pref{hamdiag} has, for given values of $N$ and $J$, a free field form, which makes it straightforward to solve the many-particle problem and in particular to determine the relevant correlation functions \cite{Haldane81}. However, for the purpose here it will be useful to reintroduce fermion variables in the expression for the Hamiltonian.

\section{Adiabatically dressed fermions}
The two sets of bosonic operators are unitarily equivalent,
\be{uni}
b_q=Ua_qU^\dag\,, b_q^\dag=Ua_q^\dag U^\dag\,,
\ee
with the unitary transformation given by
\be{utrans}
U=\exp[-\sum_{q\neq 0}{\gx_q\over 2}(a_q^2-a_q^{\dag 2})]\,.
\ee
The operator $U$ preserves the particle number of the two chiralities separately and it maps energy eigenstates of the linearized, free theory continuously into the eigenstates of the interacting theory, when the parameters $\gx_q$ are changed. With $\gw_q$ also changing smoothly, the transformation of the Hamiltonian can thus be interpreted as defining an adiabatic change from the free to the interacting theory. 

For the fermion operators the corresponding transformation is
\be{transfield}
\gf(x)=U\gy(x)U^\dag\,,
\ee
where $\psi(x)= {1\over \sqrt{L}}\sum_k e^{ikx}c_{k}$ as the original fermion operator, and $\gf(x)$ is regarded as the {\em dressed} fermion operator\cite{LeinaasHorsdal09}.   
The transformed field operator $\gf(x)$ clearly satisfies the same anticommutation relations as the original field operator $\psi(x)$, and in this sense is a fermion field. However,  the statistics of the dressed particles is not necessarily apparent in the commutation relations of the field alone,  since the form of the Hamiltonian may reveal presence of  a {\textquoteleft statistical  interactions\textquoteright} between the particles. For this reason I will examine more closely the form of the Hamiltonian, expressed in terms of dressed fermionic variables. 

To proceed the following low-energy approximation is assumed,
\be{lowen}
V_1(q)\approx V_1(0)\,,\quad V_2(q)\approx V_2(0)\,,
\ee
which for the boson frequency implies 
\be{lim2}
\gw_q\approx v_s|q|,\quad v_s=\sqrt{v_J v_N}\,.
\ee
The transformation $U$ then is approximated by
\be{utrans2}
U\approx\exp[-\sum_{q\neq 0}{\gx_0\over 2}(a_q^2-a_q^{\dag 2})]\,, \quad \tanh\xi_0=\frac{g-1}{g+1}\,,
\ee
which implies that  $U$, in this approximation, is uniquely determined by the interaction parameter $g=\sqrt{v_J/ v_N}$.

The Hamiltonian is next separated in two parts in the following way,
\be{ham3}
H=U v_s(\sum_{q\neq 0} |q| a_q^\dag a_q+{{\pi }\over{2L}}(N^2+J^2))U^\dag
+v_s{{\pi }\over{2L}}(({1\over g}-1) N^2+(g-1) J^2)\,,
\ee
where the first  term can be identified (see \pref{boseham}) as a linearized free-field Hamiltonian, with $v_s$ as Fermi velocity, and with the field variables transformed by the operator $U$. This implies that the Hamiltonian can be expressed in terms of the dressed fermion field as
\be{ham4}
H=v_s 
\{
\int_0^L dx :\sum_\chi \phi_\chi^\dag(x)(-i\chi \pd_x - k_F)\phi_\chi(x):\nn
+{\pi\over{2L}}\sum_\chi[({1\over g}+g-2)N_\chi^2
+({1\over g}-g) N_\chi N_{-\chi}]\}\,,
\ee
where the chiral fields are defined by
\be{chifield}
\psi_\chi(x)={1\over \sqrt{L}}\sum_k c_{\chi,k} e^{ikx}\,,\quad \phi_\chi(x)=U\psi_\chi(x)U^\dag
\ee
This means that, under the simplification introduced above, the Hamiltonian \pref{ham4} is unitarily equivalent to a free field Hamiltonian plus a term which is invariant under the unitary transformation.

\section{Fermi liquid description \label{four}}
The assumption of adiabatic connection between the non-interacting and interaction system forms the basis for Landau's  Fermi liquid theory. 
In this theory the total energy is given as a functional of the distribution of occupation numbers $n(k)$, associated with the non-interacting theory,
\be{efunk}
E = E[n(k)]\,, 
\ee
and the quasiparticle energies and interactions can be defined in terms of functional derivatives to first and second order in the particle density \cite{Landau56},
\be{enpart}
\gd E=\sum_k \ge(k) \gd n(k)+\half\sum_{k k'} f(k,k')\gd n(k)\gd n(k')\,.
\ee
The quasiparticles, introduced by Landau in this way, will here be identified with the dressed fermions previously discussed.
For variations about the filled Fermi sea, the expressions for energy and interactions will be referred to as  $\ge_0(k)$ and $f_0(k,k')$. 

In the present case the unitary transformation \pref{utrans2}  defines an adiabatic transition from the free field Hamiltonian to the interacting one, when the parameter $g$ is slowly changing from the initial value $1$. The expression for the variation of the energy can then be extracted directly from \pref{ham4}, and with $\chi$ in the following being restricted by the relation $\chi=\sgn \,k$, this gives 
\be{varen}
\gd E=\sum_k v_s(|k|-k_F)\gd n(k)+v_s{\pi\over L}\sum_{k,k'}(\gl_1\gq(kk')+\gl_2\gq(-kk'))\gd n(k)\gd n(k')\,,
\ee
where $\gq(k)$ is the Heaviside step function, and with $\gl_1$ and $\gl_2$ are defined by
\be{lambda}
\gl_1=\half ({1\over g}+g-2)\,,\quad \gl_2=\half ({1\over g}-g)\,. 
\ee
From this follows that the single particle energy and the interaction terms are
\be{ef}
\ge_0(k)&=&v_s(|k|-k_F)\,,\nn
f_0(k,k')&=&v_s{{2\pi}\over L}(\gl_1\gQ(kk')+\gl_2\gQ(-kk'))\,.
\ee

When corrections to the Hamiltonian \pref{ham4} are included, the above expression for the interaction can be interpreted as being valid at the Fermi points, written as
\be{int}
f_0(k_F,k_F)&=&f_0(-k_F,-k_F)=v_s{{2\pi}\over L}\gl_1\,,\nn
f_0(k_F,-k_F)&=&f_0(-k_F,k_F)=v_s{{2\pi}\over L}\gl_2\,.
\ee
The symmetric and antisymmetric combinations of the interaction terms define the two Landau parameters, which after normalization with respect to the density of states are,
\be{landau}
F_0={{L}\over{2\pi v_s}}(f_0(k_F,k_F)+f_0(k_F,-k_F))={1\over g}-1\,,\nn
F_1={{L}\over{2\pi v_s}}(f_0(k_F,k_F)-f_0(k_F,-k_F))=g-1\,.
\ee
This gives the following relation,
\be{rel}
1+F_1={1\over{1+F_0}}=g\,.
\ee

It is of interest to relate this result to the condition of Galilean invariance, as expressed in the Fermi liquid formulation. This condition is written as\cite{Landau56},
\be{gal}
\int dk \,k \,n(k)=\int dk \,m\pdd{\ge(k)}{k} n(k) \,,
\ee
where $m$ is the (bare) mass of the fermions and the occupation numbers are treated as a continuous function of $k$. In the following $n(k)$ is assumed to be the particle density normalized relative to the fully occupied system, which means that it takes values in the interval $0\leq n(k)\leq 1$.  The equation above states that the total momentum is conserved when the interaction is adiabatically turned on.
Variation in the particle density gives
\be{nvar}
\int dk\, {k\over m}\gd n(k)=\int dk \pdd{\ge(k)}{k}\gd n(k)+{L\over{2\pi}}\iint dk dk' \,\pdd{f(k,k')}{k} n(k)\gd n(k')\,,
\ee
where the last term is the result of treating $\ge(k)$ as a functional of $n(k)$.
With this being valid for arbitrary variations $\gd n(k)$, the following relation is implied,
\be{km}
{k\over m}=\pdd{\ge(k)}{k}-{L\over 2\pi}\int dk'\,f(k,k')\pdd{n(k')}{k'}\,.
\ee
For a filled Fermi sea the derivative of the particle density is
\be{dnk}
\pdd{n_0(k')}{k'}=\gd(k'+k_F)-\gd(k'-k_F)\,,
\ee
and with $k=k_F$, \pref{km} gets the form
\be{kmf}
{k_F\over m}=\left.\pdd{\ge_0(k)}{k}\right |_{k_F}+{L\over 2\pi}(f_0(k_F,k_F)-f_0(k_F,-k_F))\,.
\ee
The following relations are now  introduced,
\be{iden}
{k_F\over m}=v_F\,,\quad {k_F\over {m^*}}\equiv \left. \pdd{\ge_0(k)}{k}\right |_{k_F}=v_s\,,
\ee
with $m^*$ as the effective mass of the quasiparticles. This gives
\be{mm}
{m^*\over m}={v_F\over {v_s}}=1+F_1=g\,.
\ee
By further use of  the relations
\be{identi}
v_s=\sqrt{v_J v_N}\,,\quad g=\sqrt{v_J/v_N}\,,
\ee
the earlier result \pref{velpar} is reproduced,
\be{vj}
v_J=v_F\,,
\ee
Here this follows as consequence of Galilei invariance in Landau's Fermi liquid formulation, whereas the result in \pref{velpar} is a consequence of the corresponding symmetry of the two-particle interaction $V(k_1-k_2,q)$. The equality between $1+F_1$ and $(1+F_0)^{-1}$ in Eq.  \pref{rel}  can be seen as a consequence of the equality between the  quasiparticle velocity $k_F/m^*$ and the velocity of sound  $v_s$, as shown above.
 
\section{Fractional Statistics}
A central element in the Fermi liquid theory is the assumption that the elementary excitations (quasiparticles) of the theory obey Fermi-Dirac statistics. This means that the entropy function has the same form as for the non-interacting (bare) particles,
\be{ent}
S= -\sum_k[n(k)\ln n(k)+(1-n(k))\ln(1-n(k))]\,.
\ee
In the case discussed in the previous sections, this follows since the dressed particle field $\gf(x)$ is related to the original fermion field $\psi(x)$ by a unitary transformation. However, a further change of variable will now be introduced, which changes this relation. This is not done in the form of a transformation of the field operators, but rather by introducing a new momentum variable, with a stronger repulsion between neighboring values than demanded by the Pauli exclusion. 

Assume a set of particles occupy places in $k$-space with coordinates $k_i=2\pi n_i/L$, where $n_i$ are integers that increase monotonically with $i$. The corresponding new momentum coordinate $\gk$ is then introduced in the form of a Bethe ansatz equation,
\be{disc}
\gk_i=k_i+\gl {\pi\over L}\sum_{j\neq i}\sgn(k_i-k_j)\,,\quad i=1,2,...\,,
\ee
where $\gl$ is a new, real parameter.  
This equation leads to the following effective repulsion between the $\gk$ values,
\be{exc}
\gk_{i+1}=\gk_i +{2\pi\over L}(\gD n_i +\gl)\,,
\ee
where $\gD n_i=n_{i+1}-n_i$ is a positive integer. The equation can be interpreted as expressing that each new particle introduced in the system will occupy a one-dimensional volume $2\pi(1+\gl)/L$, as compared to $2\pi/L$ for fermions. This can be expressed more directly by the formula
\be{red}
\gD d=-(1+\gl)\gD N\,,
\ee
where $\gD d$ is the change in the number of available single-particle states within a fixed, finite interval, when $\gD N$ particles are introduced in the interval. This formulation corresponds to Haldane's defining relation of generalized exclusion statistics\cite{Haldane91}, where $d$ is interpreted as the dimension of the Hilbert space that is available for a new particle that is added to the system, and $1+\gl$ is the exclusion statistics parameter.

In the thermodynamic limit, $L\to\infty$, the relation \pref{disc} between $k_i$ and $\gk_i$ defines a mapping between the corresponding continuous variables $k$ and $\gk$, which  depend on the particle density $n(k)$ in the following way,
\be{momtrans}
\gk=k+\hf \gl\int dk'\,n(k')\,\sgn(k-k')\,.
\ee
The density $\nu(\gk)$ corresponding to the new variable $\gk$ is defined by
\be{nu}
\nu(\gk)\,d\gk=n(k)\,dk\,,
\ee
which simply states that the number of occupied states is conserved (locally) under the mapping $k\to\gk$.
It follows directly that the two densities are related by
\be{dens}
\nu(\gk)=\frac{n(k)}{1+\gl n(k)}\,,\quad n(k)=\frac{\nu(\gk)}{1-\gl \nu(\gk)}\,,
\ee
with $k$ and $\gk$ related as shown in \pref{momtrans}. With $n(k)$ limited by $0\leq n(k)\leq 1$, the corresponding restriction on $\nu(\gk)$ is $0\leq \nu(\gk)\leq 1/(1+\gl)$.

The fermion entropy \pref{ent}, in the continuum form is 
\be{entro2}
S=-{L\over{2\pi}}\int_{-\infty}^\infty dk[n(k)\ln n(k)+(1-n(k))\ln (1-n(k))]\,,
\ee
and as follows from \pref{nu} and \pref{dens}, it takes the following form when expressed in terms of the new variables,
\be{entro}
S=-{L\over{2\pi}}\int_{-\infty}^\infty d\gk[\nu(\gk)\ln\nu(\gk)-(1-\gl\nu(\gk))\ln (1-\gl\nu(\gk))\nn
+(1-(1+\gl)\nu(\gk))\ln (1-(1+\gl)\nu(\gk))]\,.
\ee
This expression agrees with expressions earlier found in
Refs.~\cite{Isakov94} and \cite{Wu94} for the entropy of exclusion statistics particles.

However, one should note that, in the present case, the transformation introduced above is so far only a change of variables. One cannot make any conclusion about the true quantum statistics of the particles without considering what the transformation makes to the energy functional of the system.   
The point to be shown is that by choosing a particular value for the parameter $\gl$, the leading part of the quasiparticle interaction, defined in the previous section by the Landau parameters $F_0$ and $F_1$, is transformed to zero. This implies that the statistics defined by the new form of the entropy is not modified by a statistical interaction term. 

The next step is then to consider how the transformation \pref{momtrans} changes the energy functions, $E[n(k)]\to E[\nu(\gk)]$, and thereby  redefines the quasiparticle energy and interaction, 
\be{enin}
\tilde\ge(\gk)={{2\pi}\over L}\frac{\gd E}{\gd\nu(\gk)}\,,\quad \tilde f(\gk,\gk')={{4\pi^2}\over L^2}\frac{\gd^2E}{\gd\nu(\gk)\gd\nu(\gk')}\,,
\ee
The idea is to express these in terms of the previous functionals $\ge(k)$ and $f(k,k')$.
For the single particle energy the transformation gives
\be{entrans}
\te(\gk')={2\pi\over L}\int dk\, \frac{\gd n(k)}{\gd\nu(\gk')}\frac{\gd E}{\gd n(k)}=\int dk\, \frac{\gd n(k)}{\gd\nu(\gk')}\ge(k)\,,
\ee
and for the interaction
\be{intfunc}
\tf(\gk'',\gk')&=&{{4\pi^2}\over L^2}\frac{\gd}{\gd\nu(\gk'')}\int dk\, \frac{\gd n(k)}{\gd\nu(\gk')}\frac{\gd E}{\gd n(k)}\nn
&=&{2\pi\over L}\int dk\,\frac{\gd^2 n(k)}{\gd\nu(\gk'')\gd\nu(\gk')}\ge(k)+\iint d\bar k\,dk 
\frac{\gd n(\bar k)}{\gd\nu(\gk'')}
\frac{\gd n(k)}{\gd\nu(\gk')} f(\brk,k)\,.
\ee
After some manipulations (see Ref.~\cite{Leinaas17}), 
the following rather simple expressions are found for the transformation matrices,
\be{transmat}
\frac{\gd n(k)}{\gd\nu(\gk')}&=&\hf\dd{}{k}[(1+\gl n(k))\sgn(k-k')]\,,\nn
\frac{\gd^2 n(k)}{\gd\nu(\gk'')\gd\nu(\gk')}&=&{1\over 4}\gl\ddtwo{}{k}[(1+\gl n(k))\sgn(k-k')\sgn(k-k'')]\,,
\ee
where the pairs of variables $k', \gk'$ and $k'', \gk''$ are related by the transformation \pref{momtrans}.

The expression obtained for the energy is then the following,
\be{entrans2}
\te(\gk')&=&\hf\int dk\, \ge(k)\dd{}{k}[(1+\gl n(k))\sgn(k-k')]\nn
&=& (1+\gl n(k'))\ge(k')+\hf \gl\int dk\, \ge(k)n'(k)\sgn(k-k')\,,
\ee
with $n'(k)=dn/dk$. 

In the case of a filled Fermi sea, the particle density and its derivative are 
\be{dens2}
n_0(k)&=&\hf(\sgn(k+k_F)-\sgn(k-k_F))\,,\nn
n'_0(k)&=&\gd(k+k_F)-\gd(k-k_F)\,.
\ee
This gives for the pseudomomentum,
\be{pseu}
\gk=k+\hf\gl\int_{-k_F}^{k_F} d\brk\,\sgn(k-\brk)=
\left\{
\begin{matrix} k+\gl k_F & \quad k>k_F \\ k(1+\gl) &\quad -k_F<k<k_F\, \,,\\ k-\gl k_F &\quad k<-k_F
\end{matrix} 
\right.
\ee
in particular $\gk_F=(1+\gl)k_F$. The transformed particle density then is
\be{trdens}
\nu_0(\gk)= {1\over{2(1+\gl)}}(\sgn(\gk+\gk_F)-\sgn(\gk-\gk_F))\,.
\ee
Introducing this in the expression for the quasiparticle energy gives
\be{etil}
\te_0(\gk)&=&
\left\{
\begin{matrix}
\quad \ge_0(\gk-\frac{\gl}{1+\gl}\gk_F)&\quad \gk>\gk_F\\[2mm]
\quad (1+\gl)\ge_0(\frac{\gk}{1+\gl})-\gl\ge_0(\frac{\gk_F}{1+\gl})&\quad |\gk|<\gk_F\quad .\\[2mm]
\quad \ge_0(\gk+\frac{\gl}{1+\gl}\gk_F)&\quad \gk<-\gk_F
\end{matrix}
\right.
\ee

For the interaction, results for the variables $k$ at the Fermi points  are only cited here (see Ref.~\cite{Leinaas17}),
\be{ftild}
\tf_0(\gk_F,\gk_F)&=&f_0(k_F,k_F)+(\gl+\hf\gl^2)\left(f_0(k_F,k_F)-f_0(k_F,-k_F)\right)+{\pi\over L}\gl^2 v_s\,,\nn
\tf_0(\gk_F,-\gk_F)&=&f_0(k_F,-k_F)-(\gl+\hf\gl^2)(f_0(k_F,k_F)-f_0(k_F,-k_F)+{2\pi\over L}v_s)\,.
\ee
If the new Landau parameters $\tilde F_0$ and $\tilde F_1$ are normalized as in \pref{landau}, this gives the following relation between these and the original Landau parameters,
\be{landpar}
\widetilde F_0&=&F_0-\gl\,,\nn
\widetilde F_1&=&(1+\gl)^2(F_1+{\gl\over {1+\gl}})\,.
\ee
Furthermore, the relation \pref{rel} between $F_0$ and $F_1$ gives the following relation between $\tilde F_0$ and $\tilde F_1$,
 \be{relnew}
 (1+\frac{\tilde F_0}{1+\gl})=( 1+\frac{\tilde F_1}{1+\gl})^{-1}\,.
 \ee
 
Assuming now that the value of the parameter $\gl$ is specified as
\be{defgl}
\gl=\gl_1+\gl_2={1\over g}-1\,,
\ee
and by using the values earlier found for $F_0$ and $F_1$ in \pref{rel}, one finds that both the new Landau parameters vanish,
\be{van}
\widetilde F_0=\widetilde F_1=0\,.
\ee
This means that both $\tf_0(\gk_F,\gk_F)$ and $\tf_0(\gk_F,-\gk_F)$ vanish, and since these functions can be shown to be continuous at the Fermi points\cite{Leinaas17}, the interactions in the low energy regime are weak, in the sense
\be{lim3}
\lim_{|\gk''|\to k_F}\,\lim_{|\gk'|\to k_F} \tf_0(\gk'',\gk')=0\,.
\ee

The conclusion is thus that the interactions of the one-dimensional fermion system effectively change the particle statistics, and make the system appear as a weakly interacting system of particles with generalized statistics. The modified exclusion parameter is given by $1+\gl=1/g$.

\section{Low energy excitations}
In the low energy approximation the interaction terms $\tilde f(\gk'',\gk')$ are negligible, and the energy function $\te(\gk)$ can be approximated by $\te_0(\gk)$, given in \pref{etil}. It is of interest to check that the sum of this energy function, over occupied particle states, reproduces the energy determined by the Hamiltonian \pref{hamdiag}.

Let us first assume the particle distribution to be without holes, with all momentum states $k_i$ being filled between a minimum value $k_{min}$, close to the Fermi point $-k_F$, and a maximum value $k_{max}$, close to $k_F$. $k_{min}$ and $k_{max}$ are then related to the particle numbers $N$ and $J$ in the following way, 
\be{nj}
k_{max}=k_F+{\pi\over L}(N+J)\,,\quad k_{min}=-k_F-{\pi\over L}(N-J)\,,
\ee
and the relation \pref{momtrans} between $\gk$ and $k$ now simplifies to
\be{simp}
\gk=(1+\gl) k-{\pi\over L}\gl J\,.
\ee
The energy contribution from the particles added to the Fermi sea, expressed in terms of the energy function $\te_0(\gk)$, is then
 \be{cont}
\gd E&=&{L\over{2\pi}}\int\te_0(\gk) \gd\nu(\gk)d\gk\nn
&=&{L\over{2\pi}}{1\over{1+\gl}}\left(\int_{\gk_{min}}^{-\gk_F}\te_0(\gk)d\gk
+\int_{\gk_F}^{\gk_{max}}\te_0(\gk)d\gk\right)\nn
&=&{L\over{2\pi}}{1\over{1+\gl}}\left(-\int_{\gk_{min}}^{-\gk_F}v_s(\gk+\gk_F)d\gk
+\int_{\gk_F}^{\gk_{max}}v_s(\gk-\gk_F)d\gk\right)\nn
&=&{\pi\over 2L} v_s((1+\gl)N^2+{1\over {1+\gl}} J^2)\,,
\ee
This expression is consistent with that of the Hamiltonian \pref{hamdiag}, with the quadratic terms in $N$ and $J$ agreeing with those in \pref{hamdiag}.
 The bosonic excitations term in \pref{hamdiag}, however, correspond to particle-hole excitations of the fermionic system, which are not included in \pref{cont}.

Particle-hole excitations can be introduced by changing the (discrete) momenta $k_i$ of the occupied states in the following way, 
\be{disk}
 k_i=k_i^0 +\gD k_i={2\pi\over L} (i+n_i)\,, \quad i=i_{min},i_{min}+1,...,i_{\max}\,.
 \ee
$k_i^0$ are the momentum values of the occupied states without holes, and  $n_i$ is a new set of integers, which introduce holes in the distribution. The condition $n_{i+1} \geq n_i $ is assumed. This makes the ordering of the set \{$k_i$\}, with respect to $i$, unchanged when the holes are introduced. For the pseudomomenta $\gk_i$ there a similar change in the values, as consequence of the transformation formula \pref{disc}, 
\be{disk2}
\gk_i&=&k_i+\gl{\pi\over L}\sum_{j=i_{min}}^{i_{max}} \sgn(k_i-k_j)\nn
&=& (2\pi/L)( i+n_i+\hf\gl\sum_{j=i_{min}}^{i_{max}}  \sgn(i-j))\nn
&\equiv&\gk_i^0+\gD k_i\,,
\ee
with $\gD k_i=(2\pi/L) n_i$. The shifts are thus the same as for the momenta $k_i$, which means that they are independent of the statistics parameter $\gl$.

With the excitations restricted to the neighborhoods of the Fermi points, linearization of the energy as function of momentum can be made, which gives
\be{nysum}
\sum_i \te_0(\gk_i)=\sum_i \te_0(\gk_i^0)+v_s\sum_i|\gD k_i|\,.
\ee
One should note that the excitation term is the same as for free fermions, although here with $v_s$ as the effective Fermi velocity. The same effect is seen in the Hamiltonian \pref{hamdiag}, when the boson frequency is linearized in $q$, $\gw_q\approx v_s |q|$. Thus, the bosonic operators $b_q^\dag b_q$ of the interacting system are unitarily equivalent with the corresponding operators $a_q^\dag a_q$ of the free system. This implies that the expression \pref{nysum} for the full energy, given as a sum over single particle energies $\te_0(\gk_i)$, reproduces the energy eigenvalues of the Hamiltonian \pref{hamdiag} within the above approximation.

\section{An application \label{seven} }
A quantum Hall system with the electrons restricted to the lowest Landau level is effectively a one-dimensional system. Thus, with $x$ and $y$ as orthogonal coordinates of electrons in the two-dimensional plane, the projection of these operators to the lowest Landau level will introduce a non-trivial commutation relation,
\be{comrel}
\com{x}{y}= -i l_B^2\,,
\ee 
with $l_B=\sqrt{1/eB}$ as the magnetic length in a strong magnetic field $B$.  This is like the commutator between coordinate and momentum in one dimension, and that is why the two dimensional $(x,y)$-space can be regarded  as the phase space of a one dimensional system. Details about how the two-dimensional description can be re-formulated as a one-dimensional description can be found in Ref.~\cite{Horsdal07}.  A short version is given here.
 
 In the Landau gauge,
 \be{}
 A_x=-yB,\quad A_y=0\,,
 \ee
 the Hamiltonian is translationally invariant in the $x$-direction and has the form of a harmonic oscillator in the $y$-direction. A complete set of (single-particle) orthonormalized wave functions is 
 \be{ortho2d}   
 \psi_k(x,y)={\cal N} e^{ikx}\psi_0(y-y_k),
 \ee
where $\cal N$ is a normalization constant, and $ \psi_0(y-y_k)$ is the ground state of the harmonic oscillator in the $y$-direction. The oscillator is $k$-dependent, with $y_k=-l_B^2k$ as the centre of the oscillator. The wavefunctions are assumed to be periodic in the $x$-direction, with period $L$. This implies that $k$ takes the discrete values $k=2\pi n/L$, with $n$ as an integer.
 In the one-dimensional (1D) description, the corresponding set of orthonormalized wave functions is given by
 \be{ortho1d}
\psi_k(\xi)={1\over \sqrt{L}}e^{ik\xi}\,.
 \ee
 The coordinate $\xi$ in 1D then corresponds to the $x$-coordinate in two dimensions (2D) while $k$ has taken the place of the $y$-coordinate.
 
 With the electrons restricted to the lowest Landau level, the Hamiltonian is completely degenerate. However, when a confining potential is introduced the degeneracy is lifted. If the potential is invariant under translation in the $x$-direction, and has the form of a harmonic oscillator in the $y$-direction, the 2D description of the single-particle Hamiltonian is
 \be{Ham2}
H&=&{1\over {2m}}(p_x+eBy)^2+{1\over {2m}}p_y^2+\half m\omega^2y^2\nn
&=&{1\over {2m}}p_y^2+\half m[(\omega_c y+{{p_x}\over{m}})^2+\omega^2y^2]\,.
\ee
Here $\omega_c=\sqrt{eB/m}$ is the cyclotron frequency of the magnetic field, $e$ is the electron charge, $m$ the mass, and $\omega$ is the angular frequency of the harmonic confining potential. The harmonic oscillator potential can be absorbed in a stronger magnetic field $\bar B$ in the following way,
\be{Ham3}
H={1\over {2m}}p_y^2+\half m\bar{\omega}_c^2(y+\frac{\omega_c}{m\bar{\omega}_c^2}p_x)^2 +{1\over {2m}}\frac{\omega^2}{\bar{\omega}_c^2}p_x^2\,,
\ee
with $\bar{\omega}_c=\sqrt{\omega_c^2+\omega^2}$ as the stronger cyclotron frequency.

The first two terms in \pref{Ham3}, which commute with the third one, is identical to the Hamiltonian of an electron in the enhanced magnetic field $\bar B$, without a confining potential. With the electron confined to the lowest Landau level of the enhanced magnetic field, the sum of the first two terms therefore define a constant, which may be disregarded. This implies that only the last term remains, and the full many-particle Hamiltonian, with a two-body interaction term $V_{ij}$ added, then has the form
\be{Ham1D}
H=\sum_i{{1}\over{2M}}p_i^2+\sum_{i<j}V_{ij}\,,
\ee
where
 \be{effmass}
M=m\,\frac{\bar{\omega}_c^2}{\omega^2}\,,
\ee
 is the effective particle mass. With $V_{ij}$ as a standard two-body interaction in 2D, the interaction in 1D is not completely local, as follows from the (small) minimum width of the electron wave functions in the lowest Landau level. In second quantized form the Hamiltonian then is
 \be{secondHam}
H=\sum_k{{\hbar^2}\over{2M}}k^2 c_k^{\dag}c_k+\frac{1}{4L}\sum_{q,k_1,k_2}V(k_1-k_2,q)c_{k_1}^{\dag}c_{k_2}^{\dag}c_{k_2-q}c_{k_1+q}\,,
\ee
where the non-locality is implicit in the dependence of $V$ on the two variables $k_1-k_2$ and $q$.
 
 This expression \pref{secondHam} for $H$ has precisely the same form as the Hamiltonian given in \pref{ham1}, here with $\ge_0(k)={{\hbar^2}\over{2M}}k^2$.
 With the ground state having the form of a filled Fermi sea, and with the low-energy states corresponding to gapless excitations, the low-energy Hamiltonian will therefore take the same form as in \pref{ham2}. As a consequence the discussion and results derived in the previous sections, will be applicable to the special case discussed here. 
 
Charge fractionalization on edges of a quantum Hall system is an interesting effect,
which has been discussed in several papers in the past \cite{LeinaasHorsdal09,FisherGlazman97,Safi97,Pham00,Trauzettel04,LeHur08,Berg09,Horsdal11,Inoue14}. The effect depends on having a system where two edges, with opposite chiralities, interact. A charge inserted on of the edges will then be separated into right-moving and left-moving components, as illustrated in Fig.~\ref{kiral}. As shown in \cite{LeinaasHorsdal09}, the right-moving components will carry a fraction ${\half(1+g)}$ of the inserted charge, and the left-moving components will carry the fraction ${\half(1-g)}$ of the charge.  This separation into right-moving and left-moving components can be explained in the Fermi liquid formulation, as will be shown in the following.

\begin{figure}[h]
\begin{center}
\includegraphics[width=7cm]{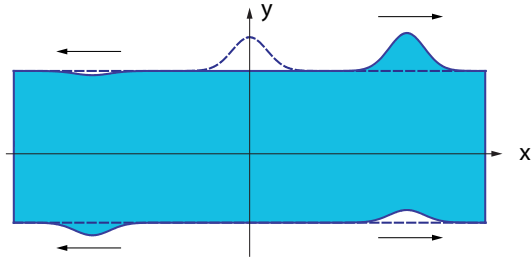}
\end{center}
\caption{\small Two-dimensional illustration of charge separation in an interacting system. A charge which is injected at the upper edge (dashed curve) is separated into a right-moving and a left-moving component due to interactions between the edges. Each of these carries a density component both on the upper and on the lower edge. The ratio between the upper and lower part of the right-going charge is determined by the interaction parameter $g$, and is the same as the ratio between the lower and upper part of the left-going charge. The initial condition restricts the total charge on the lower edge to be zero. The illustration is taken from Ref.~\cite{LeinaasHorsdal09}. \label{kiral}}
\end{figure}

Consider a situation similar to that above, with the Hamiltonian of the form \pref{secondHam}. Expressed in terms of the pseudo-momentum $\gk$, the ground state corresponds to a filled Fermi sea, with all available states in the interval $-\gk_F<\gk<\gk_F$ filled. 
The single-particle energy is assumed to have a quadratic form $\tilde\ge_0(\gk)=\hf a\gk^2$, with $a=v_s/k_F$, corresponding to the harmonic oscillator form of the confining potential in \pref{Ham2}. (In reality the precise form of the confining potential is not important for what is discussed here.)

The confining potential introduces motion of the particles to the right in the upper part of the Fermi sea and to the left in the lower part. Due to the symmetry of the occupied states in the filled Fermi sea the integrated current vanishes. To make a comparison with the situation shown in Fig.~1, a charge is next added at the upper edge and the corresponding change in the current is found and compared with the result in \cite{LeinaasHorsdal09}. However, instead of adding the charge locally, as shown in the illustration, an asymmetry is here introduced by occupying additional momentum states close to the upper edge. 

The situation is illustrated In Fig.~2. Two cases are shown, the first one with a filled Fermi sea, with $N_0=21$ particles and $N=J=0$. In the second case two particles have been added in the lowest available states close to the Fermi point $k_F$, so that $N=J=2$. The statistics parameter in the plot is chosen as $\gl=0.5$. Since there is no hole among the occupied states, there is a simple relation between the variables $k$ and $\gk$, as shown in \pref{simp},
\be{simp2}
\gk=(1+\gl) k-{\pi\over L}\gl J\,.\nonumber
\ee
By use of this relation, both $k$ and $\gk$ have been used as variables in the plots, in respectively plot (a) and (b).
Since the positions of the occupied states in plot (a) do not move along the $k$-axis when the additional charges are included, there is no obvious appearance in this plot of a separation of the additional charge into right/left-moving components. However, even if the positions of the occupied states along the $k$-axis are not changed, there is a change of positions along the energy axis, which is relevant for this separation. As shown in the figure, the {\em curve} formed by the occupied points in the two-dimensional plane is shifted along the $k$-axis when the two charges are introduced.

\begin{figure}[h]
\begin{center}
\includegraphics[width=14.5cm]{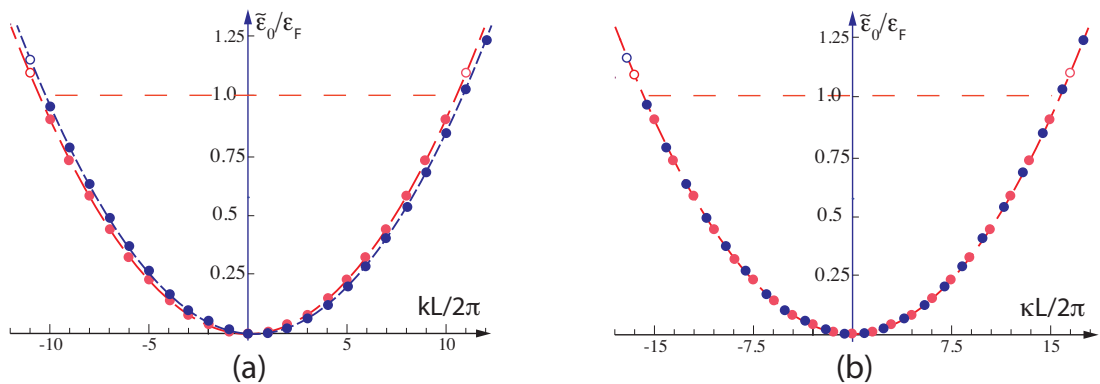}
\end{center}
\caption{\small Charge separation on edges of a quantum Hall system in a 1D representation. In $(a)$ the single particle energy $\te_0$ is shown as function of the momentum variable $k$, and in $(b)$ as a function of $\gk$.  Filled circles correspond to occupied states and open circles to unoccupied states. Two cases are shown in both plots. The first case (red circles) corresponds to a filled Fermi sea, the second case (dark blue circles) contains two additional particles, located in the lowest available momentum states close to the Fermi point $k_F$.  Figure $(b)$ shows most clearly the charge separation, with the added charge introducing a shift in the $\gk$ coordinates of the particles. This shift is interpreted as causing separation of the added charge into a right-moving component (for $\gk>0$) and a left-moving component (for $\gk<0$).  \label{energy}}
\end{figure}

In plot $(b)$, the same two cases are illustrated, here with $\gk$ rather than $k$ as variable. The charge separation is here more explicit. Thus, the two curves corresponding to $J=0$ and $J=2$ fall on the top of each other, and as a consequence the shift in energy between the two cases is linked to a shift in the variable $\gk$. This effect is a consequence of the relation between $k$ and $\gk$, shown in \pref{simp}, which implies that there is a difference between the two variables which is proportional to $J$. The interpretation now is that $\gk$ rather than $k$ determines the right/left-motion of the particles, so that positive $\gk$ corresponds to right-moving particles and negative $\gk$ corresponds to left-moving particles. With this interpretation it is straight forward to determine how the inserted charge is divided into these components, and to compare the results with what has earlier been cited from \cite{LeinaasHorsdal09}.

Assume a number $J$ of particles is added at the positive Fermi point. The shift in positions $\gk$  of the particles in Fermi sea then is
\be{}
\gD\gk=-{\pi\over L}\gl {J}=-{2\pi\over L}({1\over g}-1)J\,.
\ee
 The separation between occupied places in the Fermi sea is $\gD\gk_0={2\pi\over L}{1\over g}$, and with the combination of these expressions the result is, 
\be{} 
\frac{\gD\gk}{\gD\gk_0}=-\hf (1-g)J\,.
\ee
This can be interpreted as representing the part of the inserted charge that is transferred from the positive to the negative side of the $\gk$-axis. The result is thus that the right-moving fraction of the inserted charge is $\hf(1+g)$ and the left-moving fraction is $\hf(1-g)$, a result which agrees with what was previously found in Ref.~\cite{LeinaasHorsdal09}.

\section{Summary}
The standard approach to study interacting one-dimensional Fermi systems is to use of the bosonization technique. However, here I have focussed on how Landau's Fermi liquid theory can be applied to such systems. 

In the low-energy approximation, a unitary transformation between the excitations of the non-interacting and the interacting system is applied to rewrite the Hamiltonian in terms of {\textquoteleft dressed\textquoteright} fermion fields. 
The Hamiltonian, in this form, is then used to determine the quasiparticle energy and two-body interaction, defined as functional derivatives of the full energy of the system. With a further change of variables the interaction terms are absorbed, so that the description gets the form of a free theory. However, as shown by the transformed form of the entropy function, the quasiparticles are not fermions, but obey fractional exclusion statistics. (This means that to refer to the system as a {\em Fermi} liquid may be somewhat misleading.)

In the last section this is applied to a quantum Hall system, which is restricted to a narrow band, with interaction between the edges of the band. Although the  physical number of dimensions of the system is two, when the particles are restricted to the lowest Landau level, the dimensions are effectively reduced to one. This system is here used as an example, to demonstrate the effect of fractional exclusion statistics in the case of a one-dimensional interacting Fermi system.


\vskip5mm 
\noi


\end{document}